# Reconstruction of ionization probabilities from spatially averaged data in *N*-dimensions

**J. Strohaber\*, A. A. Kolomenskii and H. A. Schuessler**

*Texas A&M University, Department of Physics, College Station, TX 77843-4242, USA*

*\*Corresponding author: [jstroha1@physics.tamu.edu](mailto:jstroha1@physics.tamu.edu)*

We present an analytical inversion technique which can be used to recover ionization probabilities from spatially averaged data in an *N*-dimensional detection scheme. The solution is given as a power series in intensity. For this reason, we call this technique a multiphoton expansion (MPE). The MPE formalism was verified with an exactly solvable inversion problem in 2D, and probabilities in the postsaturation region, where the intensity-selective scanning approach breaks down, were recovered. In 3D, ionization probabilities of Xe were successfully recovered with MPE from simulated (using the ADK tunneling theory) ion yields. Finally, we tested our approach with intensity-resolved benzene ion yields showing a resonant multiphoton ionization process. By applying MPE to this data (which was artificially averaged) the resonant structure was recovered—suggesting that the resonance in benzene may have been observable in spatially averaged data taken elsewhere.

In the investigation of fundamental quantum mechanical phenomena, the interaction of ultrashort pulses of radiation with atomic and molecular systems has become an important subject in the laboratory and theoretical investigations. In particular, the interaction of the noble gases with radiation has been studied extensively over the past few decades and is well documented in the literature [1—3]. With advances in theory and technology, a refined physical understanding of the ionization of the rare-gas atoms is currently being elaborated [1, 4]. In the case of complex molecular systems, the task of calculating ionization probabilities becomes increasingly more difficult as the number of degrees of freedom increases. For this reason approximations designed to capture the essential physics of the interaction are routinely proposed [5,6], but only through experimental investigations can these approximations be verified and potentially lead to useful generalizations.

In many experiments, measured data is the result of integration by a detection device. For example, in the interaction of radiation with ion beams, 3D velocity distributions of product particles (*i.e.,* photofragments and photoelectrons) are projected onto a 2D detector [7—9]. In the production of plasma channels, 3D radial electron density profiles are projected onto a 2D surface and recorded as interferograms [10,11]. These are examples where the Abel inversion has been used to render physical information from integrated data. Analogously, time-of-flight spectrometers commonly integrate 3D ion distributions produced within the focus of a laser beam [12]; consequently, measured yields are averaged over a broad range of intensities. This is known in the literature as spatial averaging [13].

To understand spatial averaging and some of its consequences, consider the isointensity shells within a $HG_{0,0}$ mode. At a particular intensity $I_j$, the boundary of an isointensity shell has the dependence $r_j(z,I_0) = w\sqrt{\ln\left|I_0 w_0^2 / I_j w^2\right|/2}$, where $w_0$ and $w(z)$ are the beam waist and size, and $I_0$ is the peak intensity. As an example, in Fig. 1 the blue peanut-shaped shell, having a semitransparent top-half, is calculated for a lower intensity than the red shell shown within. If we assume that the appearance and saturation intensities of $Xe^+$ correspond to the blue and red shells respectively, and the saturation intensity of $Xe^{2+}$ coincides with the peak intensity $I_0$, then $Xe^+$ ions will be mainly found within the volume bounded by blue and red shells, while those of $Xe^{2+}$ within the red shell. For this reason, experimentally measured ion signals, for peak intensity $I_0$, are the volume-integrated product yield [12,13].

$$I_0 S(I_0) \propto \int_0^{I_0} I_0 \left| \frac{\partial V(I, I_0)}{\partial I} \right| P(I) dI \tag{1}$$

Here $S(I_0)$ is the ion signal, $P(I)$ is the intensity-dependent ionization probability, and $\left| \partial V(I, I_0) / \partial I \right|$ is the so-called volumetric weighting factor. For reasons that will become clear, we have multiplied both sides of Eq. (1) by $I_0$. The physical manifestation of Eq. (1) has caused many difficulties for both experimentalists and theoreticians. In experiments, the effect of spatial averaging tends to smear out subtle features that may have otherwise appeared in presaturation yields and removes expected decreases (typically orders of magnitude) in postsaturation yields due to competing higher-order processes such as sequential ionization or fragmentation. In the literature, with some exceptions [4,14], theoretically determined ionization probabilities are artificially averaged—possibly concealing new physics—for better comparison with experimental data. In this communication, an analytical solution to Eq. (1), which can be used to recover $P(I)$ from $S(I_0)$ in an $N$-dimensional detection scheme, is derived. Previously experimental and theoretical methods used to circumvent spatial averaging are: intensity-

selective scanning (ISS) [13] and its modifications [4], intensity-difference spectrum [15], and intensity resolved ion imaging (Ref. [12] and references therein).

For the general *N*-dimensional problem, the highly successful methods outlined in [4, 13,15] cannot be used to recover $P(I)$. The volumes of the isointensity shells in *N*-dimensions are [15]:

$$V_{3D} \propto \left(\frac{I_0}{I}-1\right)^{1/2} + \frac{1}{6}\left(\frac{I_0}{I}-1\right)^{3/2} - \arctan\left(\frac{I_0}{I}-1\right)^{1/2},  \qquad (2a)$$

$$V_{2D} \propto \ln\left|\frac{I_0}{I}\right|,  \qquad (2b)$$

$$V_{1D} \propto \left(\ln\left|\frac{I_0}{I}\right|\right)^{1/2}.  \qquad (2c)$$

Each of the volumes in Eq. (2) is dependent on both the local and peak intensities $V_{ND} = V_{ND}(I, I_0)$. By using a sufficiently large entrance slit to a TOF spectrometer, all ions in the focus of a laser beam can be collected. This type of detection method is known as full view and corresponds to the volume in Eq. (2a) [13]. The volume in Eq. (2b) is that of a 2D slice taken perpendicular to the propagation direction, which is achieved by using a narrow rectangular slit in the *xy*-plane (Fig. 1). Two different detection schemes are possible with this geometry: intensity scanning (IS) and ISS. The inversion technique that we outline here is applicable to both types of scanning. The 1D volume of Eq. (2c) is that of a line-volume along the propagation direction of the beam. This 1D volume can be realized experimentally by placing a narrow slit along the propagation direction (*z*), and in the *y*-direction the volume can be further limited by time slicing [12,16,17]. Taking the derivatives of the volumes in Eq. (2) and multiplying by $I_0$ gives the intensity-scaled volumetric weighting factors $K_{ND} = I_0 |\partial V / \partial I|$ in Eq. (1):

$$K_{3D}(I,I_0) \propto \frac{I_0}{I}\left(1+\frac{I_0}{I}\right)\left(\frac{I_0}{I}-1\right)^{1/2}, \qquad (3a)$$

$$K_{2D}(I,I_0) \propto \frac{I_0}{I}, \qquad (3b)$$

$$K_{1D}(I,I_0) \propto \frac{I_0}{I}\left(\ln|I_0/I|\right)^{-1/2}. \qquad (3c)$$

To solve for the probability, we make the assumption that both $P(I)$ and the intensity-scaled signal $I_0 S(I_0)$ can be expanded in series:

$$I_0 S(I_0) = I_0^m \sum_k A_k f_k(I_0), \qquad (4a)$$

$$P(I) = I^n \sum_k B_k f_k(I). \qquad (4b)$$

Here $A_k$ and $B_k$ are expansion coefficients, and $f_k(I_0)$ are basis functions. The role of the integers $m$ and $n$ will be discussed later. If $P(I)$ can be presented analytically, and $B_k$ and $f_k(I)$ are known, then the probability can be expanded using Eq. (4b). To this end, Eqs. (4) are inserted into the volume integral of Eq. (1)

$$I_0^m \sum_k A_k f_k(I_0) \propto \sum_k \int_0^{I_0} K_{ND}(I,I_0) I^n B_k f_k(I) dI. \qquad (5)$$

Here, the summation has been moved outside of the integral. From Eq. (3) it can be seen that all of the kernels have a commonality: these kernels can be written in the form $K_{ND}(I, I_0) = K_{ND}(I/I_0)$. This explains why both sides of the volume integral of Eq. (1) were multiplied by $I_0$, and allows the substitution $\xi = I/I_0$ to be made to all intensity-dependent quantities on the right-hand side of Eq. (5),

$$I_0^m \sum_k A_k f_k(I_0) \propto I_0^{n+1} \sum_k \int_0^1 K_{ND}(\xi) \xi^n B_k f_k(\xi I_0) d\xi. \tag{6}$$

Both sides of Eq. (6) will have a similar form if $n+1 = m$, $B_k = A_k / G_k$, and $f_k(\xi I_0) = f_k(\xi) f_k(I_0)$ (the last equality is Cauchy's multiplicative equation [18])

$$I_0^m \sum_k A_k f_k(I_0) \propto I_0^m \sum_k A_k f_k(I_0) \left[ \frac{1}{G_k} \int_0^1 K_{ND}(\xi) \xi^{m-1} f_k(\xi) d\xi \right]. \tag{7}$$

For both sides of Eq. (7) to be consistent, the bracketed expression together with the proportionality constant must be equal to unity for all values of $k$ and $m$, therefore

$$G_k \propto \int_0^1 K_{ND}(\xi) \xi^{m-1} f_k(\xi) d\xi. \tag{8}$$

Equation 5 has now been reduced to a form that allows the values of $B_k = A_k / G_k$ to be determined if those of $A_k$ are known. Because the basis functions $f_k$ satisfy Cauchy's equation, they have the solutions $f_k(x) = x^k$. The series solutions of Eqs. (4) take the final form

$$S(I_0) = I_0^{m-1} \sum_k A_k I_0^k, \tag{9a}$$

$$P(I) = I^{m-1} \sum_k \frac{A_k}{G_k} I^k . \tag{9b}$$

The *volumetric coefficients* $G_k$, which inherit the *geometry* of the problem at hand and are needed in order to recover $P(I)$, can be found analytically by inserting Eqs. (3) into Eq. (8) with $f_k(\xi) = \xi^k$ and integrating:

$$G_k^{3D} \propto \frac{(m+k-2)^2 \left(2(m+k-3)\right)!}{4^{m+k} \left((m+k-2)!\right)^2 (m+k-1)}, \tag{10a}$$

$$G_k^{2D} \propto \frac{1}{m+k-1}, \tag{10b}$$

$$G_k^{1D} \propto \sqrt{\frac{1}{m+k-1}} . \tag{10c}$$

To better understand Eq. (9b), consider a MPI process $P(I) \propto I^\alpha$ of arbitrary order $\alpha$ in 2D. Integrating this $P(I)$ in Eq. (1) with Eq. (3b) gives $S(I_0) \propto I_0^\alpha / \alpha$. Here $A = 1/\alpha$ is due to spatial averaging. Using Eq. (8) we find that $G = 1/\alpha$, so that $B = 1$; thus, recovering the probability. While the physics of the ionization process is in general more complex than MPI alone (i.e., tunneling and over the barrier ionization may contribute), we find it convenient to think of $P(I)$ in Eq. (9b) as a multiphoton expansion (MPE).

Despite the lengthy derivation, recovering $P(I)$ from $S(I_0)$ is relatively straightforward. In this work, ion yields having roughly 60 data points (a typical experimental data set) were splined in order to increase the number of points by an order-of-magnitude. The splined data was then expanded in a polynomial series to determine the values of $A_k$. The volumetric coefficients

$G_k$, provided by Eq. (10), along with the numerically determined values of $A_k$ were then used to recover $P(I)$ from Eq. (9b).

As a first demonstration of MPE, which can be readily verified by the reader, an exactly solvable inversion problem in 2D is carried out using the model probability

$$P(I) = \frac{(I/I_{S1})^{n_1}}{1+(I/I_{S1})^{n_1}} - \frac{(I/I_{S2})^{n_2}}{1+(I/I_{S2})^{n_2}}. \tag{11}$$

Here $n_1$ and $n_2$ are MPI orders, and $I_{S1}$ and $I_{S2}$ are saturation intensities chosen to coincide with those of $Xe^+$ and $Xe^{2+}$ ionized by 800 nm, 100 fs radiation ($n_1 = 8$, $n_2 = 14$, $I_{S1} \sim 8\times 10^{13}$ W/cm$^2$ and $I_{S2} \sim 2\times 10^{14}$ W/cm$^2$). The spatially averaged yield $S(I_0) = \ln\left|(I/I_{S1})^{n_1}+1\right|/n_1 - \ln\left|(I/I_{S2})^{n_2}+1\right|/n_2$ was found by inserting Eq. (11) and Eq. (3b) into Eq. (1) and integrating. Using the formalism of MPE, the probability was found to be

$$P(I) = \sum_k (-1)^{k+1}\left[\left(\frac{I}{I_{S1}}\right)^{kn_1} - \left(\frac{I}{I_{S2}}\right)^{kn_2}\right]. \tag{12}$$

The series in Eq. (12) is precisely the series expansion of our model probability Eq. (11) and is plotted (solid curve) in Fig. 2(a) along with its spatially averaged yield $S(I_0)$ (dashed curve). The blue circles are the recovered probabilities obtained by *numerically* fitting to $I_0^{1-m}S(I_0)$ [Eq. (9a)] and expanding $P(I)$ with Eq. (9b) ($m = 0$ and $k = 36$). For comparison, we have also plotted the recovered probability using the ISS approach (red squares) [13]. As pointed out in Ref. [4], ISS breaks down after saturation. In contrast, the recovered probability using MPE works for the entire yield curve.

The effects of $I^m$ for various values of $m$ in 2D are demonstrated graphically in Fig. 2(b). For each set of data, the order of the expansion was held constant $k=10$: red squares ($m=8$) and blue circles ($m=-8$). In 2D, the values of $m$ are not restricted Eq. (10b). In contrast, *integer* values of $m$ for the 3D and 1D cases are restricted to $m \geq 3$ and $m \geq 1$ respectively, Eq. (10a) and Eq. (10c). The restrictions on $m$ mean that in 3D MPE is applicable to MPI process of order $\geq 2$ and for all orders in 1D and 2D. Additionally, the recovered probabilities in Fig. 2(b) (red squares) imply that fitting of $P(I)$ in 3D and 1D will favor presaturation yields because of the restrictions on $m$. Better fitting to postsaturation yields have been achieved by increasing the order of the expansion, weighting data in this region or by treating both regions separately.

As a first example in 3D, theoretically calculated ionization probabilities of $Xe^+$ and $Xe^{2+}$, using ADK [3], were spatially averaged and inverted using MPE ($m=3$ and $k=36$). Excellent agreement was found between the theoretical (black curves) and recovered probabilities [blue squares and red circles, Fig. 3(a)]. As our final example in 3D, yields of experimentally obtained benzene parent molecular ions (to be published elsewhere), obtained with the imaging spectrometer of [12], are shown as the black circles in Fig 3(b). Because ions measured in this fashion are collected from a region of nearly constant intensity, the data is expected to *resemble* true ionization probabilities; therefore, we treat this data as $P(I)$. To recover the probabilities from the *artificially* averaged data (dashed curve), we set $m=3$ and $k=36$. The recovered probability (blue squares) is in agreement with the measured "probability" reproducing structures on the leading edge of the curve, which have been attributed to resonant multiphoton ionization through the $S_1$ intermediate state. This demonstration suggests that not only could this previously unseen resonant multiphoton process have been observed in data taken elsewhere ([19] and references therein), but previous experimental data collected for other systems could also be reanalyzed using MPE. The red triangles have been obtained by weighting the postsaturation signal over the presaturation signal to favor fitting after saturation. The red triangles reproduce the postsaturation decrease and the overall structure of the probability. They do not, however, reproduce the finer structures on the leading edge of the curve, which is understandable because less significance has been placed on this data.

In our simulations, we have typically found that the recovered probability converges to the input probability or diverges with noticeable oscillations [i.e., the red squares and blue circles in Fig. 2(b)]. If the postsaturation probabilities cannot be recovered or appear to be unreliable, then $m$ can be set to a larger positive value (favoring fitting to presaturation yields), and the method of conserved probability of ionization [4] may be applied. In this case, MPE take an analogous role in 3D that ISS plays in 2D. As a final note, as with most inverse techniques, the signal-to-noise ratio plays an important part in the inversion process [13]. An advantage of MPE in 3D is the accumulation of larger amounts of data compared to geometries of lesser dimensionality.

To the best of our knowledge, we are the first to demonstrate an analytical inversion technique to recover ionization probabilities from spatially averaged data in an $N$-dimensional detection scheme. In contrast to ISS, our 2D inversion technique is capable of recovering postsaturation probabilities. Finally, we are currently constructing an imaging spectrometer of the type demonstrated in Ref. [12] to investigate stabilization in atomic and molecular systems. This spectrometer will be able to operate in either full-view or imaging mode, allowing for further testing of MPE by comparing data taken by each mode of operation.

This work was partially supported by the Robert A. Welch Foundation (grant No. A1546), the National Science Foundation (grants Nos. 0722800 and 0555568), the Air Force Office of Scientific Research (grant FA9550-07-1-0069) and Qatar National Research Fund (grant NPRP30-6-7-35).

**Figure Captions**

Fig. 1 (color online). Isointensity shells within a Gaussian laser beam propagating along the $z$-axis. The blue shell, having semi-transparent top-half, is at a lower intensity than the red shell shown within.

Fig. 2 (color online). Reconstruction of a model probability in 2D: (a) model probability (black curve, see text), spatially averaged probability (dashed curve), recovered probabilities using MPE (blue circles), and recovered probabilities using ISS (red squares; the plot is shifted down for better viewing); (b) recovered probabilities using MPE from the spatially averaged probability in (a) for $m = 8$ (red squares) and $m = -8$ (blue circles).

Fig. 3 (color online). Reconstruction of xenon ionization probabilities calculated with ADK theory in 3D: (a) calculated (black solid curves) and recovered (blue squares and red circles) probabilities of $Xe^+$ and $Xe^{2+}$; the spatially averaged probability is shown by the dashed curve. (b) Ion yields of the benzene parent molecule obtained using an imaging spectrometer (black circles): artificially averaged benzene yield (dashed curve), recovered probability using MPE (blue squares) and recovered probability obtained by weighting the postsaturation ionization yields (red triangles, see text).

# Figures

Figure 1

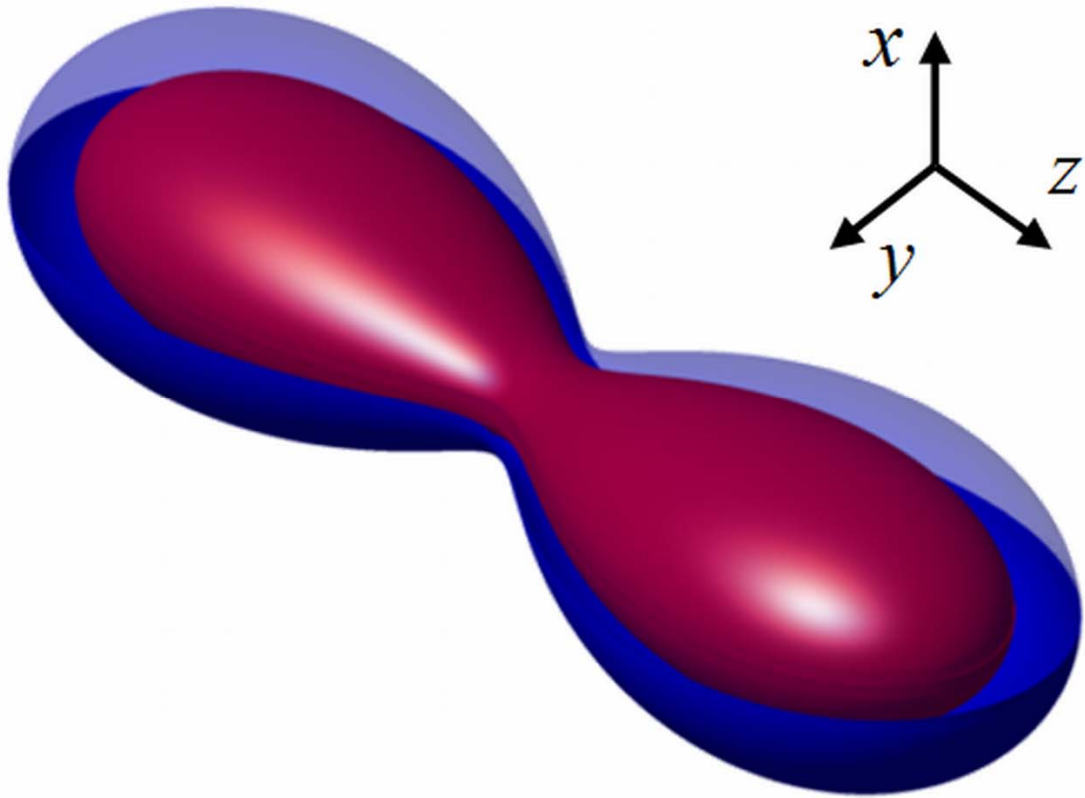

Figure 2

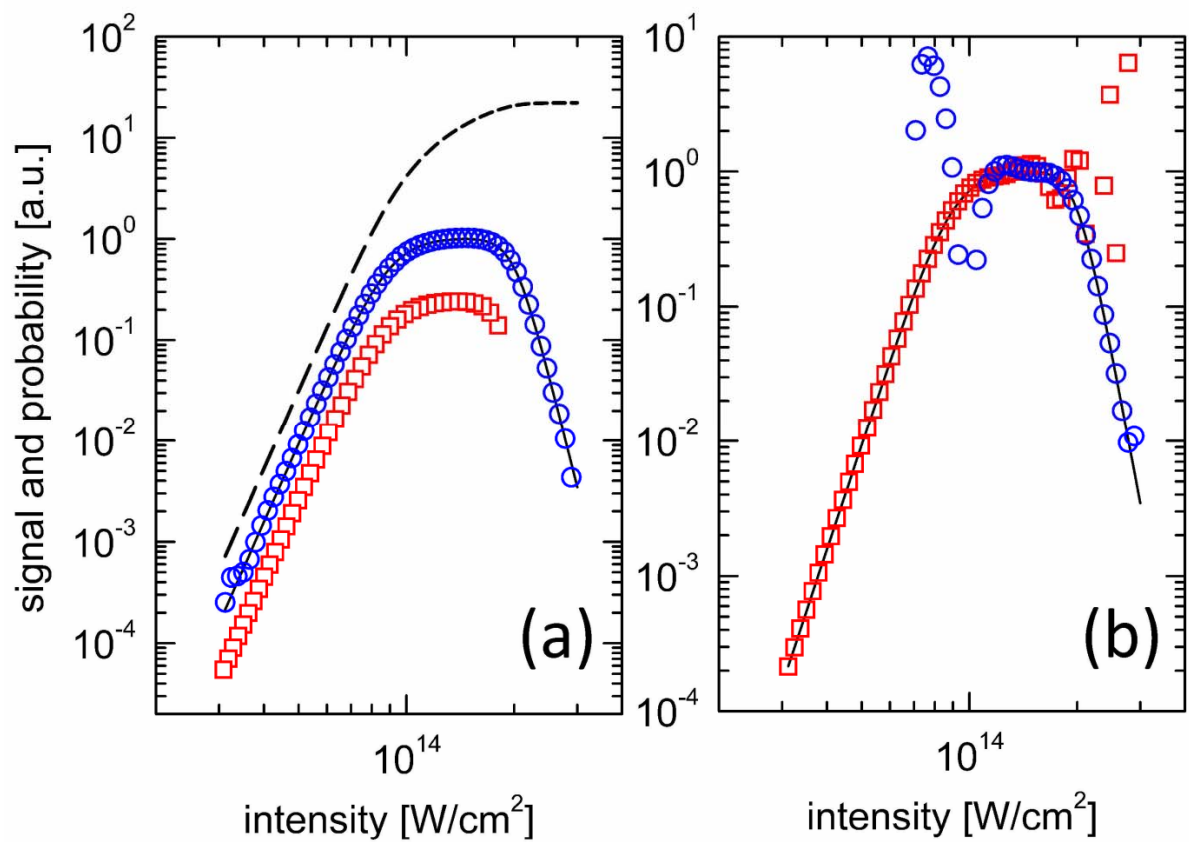

Figure 3

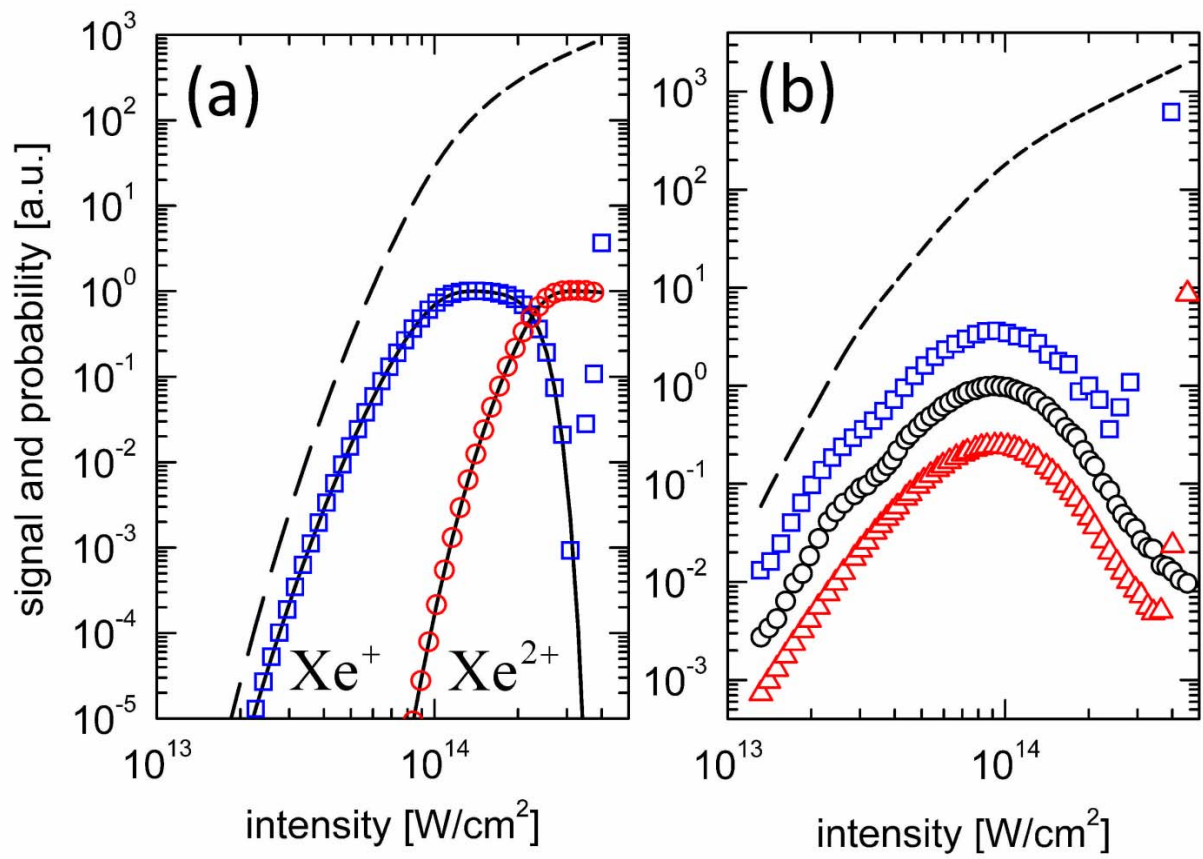